\documentclass[a4paper,aps,prd,10pt,preprintnumbers,showpacs,twocolumn,superscriptaddress,nofootinbib,amsmath,amssymb]{revtex4-1}
\usepackage{amssymb}
\usepackage{amsfonts}
\usepackage{amsmath}
\usepackage{graphicx}
\usepackage{dcolumn}
\usepackage{bm}
\usepackage{makeidx}
\usepackage[latin1]{inputenc}
\usepackage[english, english]{babel}
\usepackage[dvips]{hyperref}
\usepackage{color}
\usepackage{booktabs}

\usepackage{graphicx}
\def\be{\begin{equation}}
 \def\ee{\end{equation}}
 \def\bea{\begin{eqnarray}}
 \def\eea{\end{eqnarray}}

\def\aa{\widetilde{\alpha}}

\makeindex

\begin{document}

\title{Quasinormal modes of a scalar field in the Einstein--Gauss--Bonnet-AdS black hole background: Perturbative and nonperturbative branches}
\author{P. A. Gonz\'{a}lez}
\email{pablo.gonzalez@udp.cl}
\affiliation{Facultad de Ingenier\'{\i}a y Ciencias, Universidad Diego Portales, Avenida Ej\'{e}%
rcito Libertador 441, Casilla 298-V, Santiago, Chile}

\author{R. A. Konoplya}
\email{roman.konoplya@gmail.com}
\affiliation{Theoretical Astrophysics, Eberhard-Karls University of T\"ubingen, T\"ubingen 72076, Germany}

\author{Yerko V\'{a}squez}
\email{yvasquez@userena.cl}
\affiliation{Departamento de F\'{\i}sica y Astronom\'ia, Facultad de Ciencias, Universidad de La Serena,\\
Avenida Cisternas 1200, La Serena, Chile}
\date{\today}

\begin{abstract}

It has recently been found that quasinormal modes of asymptotically  anti-de Sitter (AdS) black holes in theories with higher curvature corrections may help to describe the regime of intermediate 't Hooft coupling in the dual field theory. Here, we consider quasinormal modes of a scalar field in the background of spherical Gauss--Bonnet--anti-de Sitter (AdS) black holes. In general, the eigenvalues of wave equations are found here numerically, but at a fixed Gauss-Bonnet constant $\alpha = R^2/2$  (where $R$ is the AdS radius), an exact solution of the scalar field equation has been obtained. Remarkably, the purely imaginary modes, which are usually appropriate only to some gravitational perturbations, were found here even for a test scalar field. These purely imaginary modes of the Einstein--Gauss--Bonnet theory do not have the Einsteinian limits, because their damping rates grow, when $\alpha$ is decreasing. Thus, these modes are nonperturbative in $\alpha$. The real oscillation frequencies of the perturbative branch are linearly related to their Schwarzschild-AdS limits $Re (\omega_{GB}) = Re (\omega_{SAdS}) (1+ K(D) (\alpha/R^2))$, where $D$ is the number of spacetime dimensions. Comparison of the analytical formula with the frequencies found by the shooting method allows us to test the latter. In addition, we found exact solutions to the master equations for gravitational perturbations at $\alpha=R^2/2$ and observed that for the scalar type of gravitational perturbations an eikonal instability develops.

\end{abstract}

\maketitle




\section{Introduction}

Perturbations and proper (quasinormal) oscillations of black holes has been an intensively developing topic during the past 15 years \cite{Konoplya:2011qq}. The great impetus has recently been done by the observation of gravitational waves from, apparently, a merger of two black holes  \cite{Abbott:2016blz}. Although the observed signal is consistent with the Einstein gravity \cite{TheLIGOScientific:2016src}, the window for alternative theories is also open \cite{Konoplya:2016pmh}, owing to large uncertainties in the determination of mass and angular momenta of the ringing black hole. The higher curvature corrections to the Einstein gravity, given in the form of the second order in the curvature (Gauss-Bonnet) term, is one of the most interesting alternatives because they are predicted by the low-energy limit of string theory.

Quasinormal modes of asymptotically anti-de Sitter (AdS) black holes play crucial role in the holographic description of quark-gluon plasmas. In Ref. \cite{Kovtun:2004de}, it was shown that for various gravitational backgrounds the holography predicts the universal upper limit for strongly coupled systems in the conformal field theory,
\begin{equation}\label{qq}
\frac{\eta}{s} \approx \frac{\hbar}{4 \pi k},
\end{equation}
where $\eta$ is the shear viscosity and $s$ is volume density of entropy. Soon this theoretical prediction (\ref{qq}) was confirmed when observing quark-gluon plasma at the Relativistic Heavy Ion Collider \cite{Luzum:2008cw}.

The essential point of AdS/CFT-inspired calculations, aimed at the description of quark-gluon plasmas, is that the 't Hooft coupling $\lambda$ is implied to be large. At small coupling $\lambda$, one can describe the system in terms of the kinetic theory. The regime, unknown up to now, is a transition from strong to weak coupling, which is necessary if we want to have the full and reliable theoretical description. Recently, attempts to find the approach to the intermediate coupling regime have been made through the analysis of gravitational theories with higher derivatives, such as Gauss-Bonnet (GB), Lovelock, $R^4$, and others  \cite{Waeber:2015oka}, \cite{Grozdanov:2016fkt}, \cite{Grozdanov:2016vgg}.

There are a few papers devoted to numerical analysis of quasinormal modes of Gauss-Bonnet and Lovelock black holes in asymptotically flat and de Sitter spacetimes \cite{Konoplya:2004xx}, \cite{Cuyubamba:2016cug}, \cite{Yoshida:2015vua}, \cite{Konoplya:2017wot}, \cite{Chen:2016fuy}. The gravitational quasinormal modes and the dual hydrodynamic regime for Gauss--Bonnet--anti-de Sitter black holes with the planar horizon (i. e., black branes)  were analyzed in Ref. \cite{Grozdanov:2016fkt}, \cite{Grozdanov:2016vgg}.  The gravitational modes for the corresponding spherical black holes were numerically found in Ref. \cite{Konoplya:2017ymp} for Gauss--Bonnet theory and in Ref. \cite{Konoplya:2017lhs} for the generic Lovelock theory, where it was also shown that for some range of parameters (GB coupling $\alpha$ and  black hole radius $r_H$) black holes are unstable. The found gravitational instability is ``driven'' by a new branch of modes, which are nonperturbative in $\alpha$ (that is, they do not exist in the limit $\alpha =0$). This, eikonal, instability is similar to the instability found  for spherical asymptotically flat \cite{Gleiser:2005ra}, \cite{Dotti:2005sq}, \cite{Takahashi:2010gz} and planar AdS \cite{Takahashi:2011du} black holes. The instability is accompanied by the breakdown of the well-posedness of the initial values problem \cite{Reall:2014pwa}. A single, rather isolated case is a five-dimensional black hole at $\alpha =R^2/2$, which, although unstable \cite{Konoplya:2017ymp}, is worth investigating,  because the metric becomes greatly simplified, so that one can find an analytic solution of the perturbation equation. There are not many examples of exact solutions for quasinormal modes of black holes, and most of them are in the lower-than-$3+1$ -dimensional spacetimes \cite{exactQNM}, \cite{Grozdanov:2016fkt}. Exact quasinormal frequencies of a test scalar field were found in Ref. \cite{Gonzalez:2010vv}  for the Chern-Simons black holes. Unlike numerical data, the analytical formula for quasinormal modes makes it easier to understand the nature of the new modes and helps to check the correctness and accuracy of the numerical techniques used earlier.

As the test scalar field is known to be free from the eikonal instability at least in the asymptotically flat spacetime \cite{Konoplya:2004xx} and the new, purely imaginary modes are related to this instability \cite{Konoplya:2017ymp}, it is not easy to predict whether such  nonperturbative modes exist also for test scalar field perturbations. At the same time, it is interesting to know whether the existence of a nonperturbative branch depends on the spin of a field under consideration. When $\alpha =0$, a scalar field in the background of the Schwarzschild-AdS black holes  does not have purely imaginary modes in its spectrum \cite{Kuang:2017cgt, Horowitz:1999jd, Konoplya:2002ky}.

Having in mind the above motivation, we shall calculate quasinormal modes of a scalar field in the background of the $D=5, 6$ -dimensional Einstein--Gauss--Bonnet-AdS black hole. We shall show that for those values of $\alpha$ which correspond to stable black holes \cite{Konoplya:2017ymp} the spectrum of scalar quasinormal modes consists of two branches. One is \emph{perturbative} in $\alpha$ and goes over into the known Schwarzschild-AdS quasinormal modes \cite{Horowitz:1999jd}. The other branch is nonperturbative: when $\alpha$ is decreasing, the damping rate of (purely imaginary) quasinormal modes increases and does not go over into the Schwarzschild-AdS modes.

In the $D=5$ black hole spacetime at a fixed Gauss-Bonnet coupling constant $\alpha =R^2/2$, we find exact solutions for the gravitational and test scalar field perturbation equations at the Dirichlet and Neumann (for a scalar field allowing for tachyons) boundary conditions. It turns out that the scalar field is unstable under Neumann boundary conditions. We shall demonstrate analytically the appearance of the eikonal instability and $\ell$ divergence for gravitational perturbations in the case $\alpha =R^2/2$.
The numerical shooting we used turned out to be difficult to apply effectively in the regimes of higher overtones, small black hole radii, or higher multipoles.

The paper is organized as follows. Section II briefly relates the essentials of the Gauss-Bonnet black hole background. Section III gives basic information on perturbation equations and discusses in detail numerical data on quasinormal modes for a test scalar field at various values of $\alpha$ in the region of stability. In addition, we compare the numerical results with the obtained analytically formulas. Section IV is devoted to the analytical deduction of the exact solutions for the perturbation equation in the form of hypergeometric functions.  In Section V, we review the obtained results and mention the future prospects.

\section{Einstein--Gauss--Bonnet-AdS black holes}
\label{Background}

The Lagrangian of the $D$-dimensional Einstein-Gauss-Bonnet theory is given by the relation
\begin{equation}\label{gbg3}
  \mathcal{L}=-2\Lambda+R+\frac{\alpha}{2}(R_{\mu\nu\lambda\sigma}R^{\mu\nu\lambda\sigma}-4\,R_{\mu\nu}R^{\mu\nu}+R^2).
\end{equation}
An exact solution for a static spherically symmetric black hole in the $D$-dimensional Einstein-Gauss-Bonnet theory (\ref{gbg3}) has the form \cite{Boulware:1985wk}
\begin{equation}\label{gbg4}
 ds^2=-f(r)dt^2+\frac{1}{f(r)}dr^2 + r^2\,d\Omega_n^2,
\end{equation}
where $d\Omega_n^2$ is a $(n=D-2)$-dimensional sphere and
$f(r)=1-r^2\,\psi(r)$,
such that it satisfies the following relation:
\begin{equation}\label{Wdef}
W[\psi]\equiv\frac{n}{2}\psi(1 + \aa\psi) - \frac{\Lambda}{n + 1} = \frac{\mu}{r^{n + 1}}\,.
\end{equation}
Further properties of this solution were analyzed in Ref. \cite{Cai:2001dz}. Here, the Gauss--Bonnet coupling constant $\aa$ is
$\aa \equiv \alpha (n - 1) (n - 2)/2,$
and $\mu$ is a constant, proportional to mass. The  solution of Eq. (\ref{Wdef}), which goes over into the known Tangherlini solutions \cite{Tangherlini:1963bw}, allowing for a nonzero $\Lambda$ term, is
\begin{equation}\label{psidef}
  \psi(r)=\frac{4\left(\frac{\mu}{r^{n+1}}+\frac{\Lambda}{n+1}\right)}{n+\sqrt{n^2+8\aa n\left(\frac{\mu}{r^{n+1}}+\frac{\Lambda}{n+1}\right)}}.
\end{equation}
We are interested in this branch of solutions because it has the known Einsteinian ($\alpha =0$) asymptotically flat, de Sitter and anti-de Sitter limits. When $\Lambda =0$, there is another branch of asymptotically anti-de Sitter solutions, which does not have the Einsteinian limit.

To measure all quantities in the units of the same dimension, we express $\mu$ as a function of the event horizon $r_H$ as \cite{Cuyubamba:2016cug}
\begin{equation}\label{massdef}
  \mu=\frac{n\,r_H^{n-1}}{2}\left(1+\frac{\aa}{r_H^2}-\frac{2\Lambda  r_H^2}{n(n+1)}\right).
\end{equation}
We shall measure $\Lambda$ in units of the AdS radius $R$ [defined by relation $\psi(r\rightarrow\infty)=-1/R^{2}$]. Then,
\begin{equation}\label{AdSlambda}
  \Lambda=-\frac{n(n+1)}{2R^2}\left(1-\frac{\aa}{R^2}\right),
\end{equation}
implying that $\aa<R^2$. In the $D=5$ case, $\alpha =\aa$, and when, in addition, $\alpha = R^2/2 $; then, the metric function $f(r)$ has a Ba\~nados--Teitelboim--Zanelli (BTZ)-like \cite{Banados:1992wn} form,
\begin{equation}\label{metric}
f(r)=\frac{r^2}{R^2}+1-\sqrt{\frac{4\mu}{3R^2}}=\frac{r^2-r_H^2}{R^2}.
\end{equation}


\section{Quasinormal modes}

Here, we shall consider perturbations of a test scalar field for various values of $\alpha$. As gravitational perturbations have recently been considered in Ref. \cite{Konoplya:2017ymp}, here, we analyze gravitational perturbations only  of the $\alpha = R^2/2$ case, which can be treated analytically.

Perturbations of a test scalar field obey the general relativistic Klein-Gordon equation
\begin{equation}
\frac{1}{\sqrt{-g}}\partial _{\mu }\left( \sqrt{-g}g^{\mu \nu }\partial
_{\nu } \varphi \right) =m^{2}\varphi \,, \label{KGNM}
\end{equation}%
where $m$ is the mass of the scalar field $\varphi $. With the help of the following ansatz
$\varphi =e^{-i\omega t}Y(\Omega)R(r),$
the Klein-Gordon equation reduces to the form
$$  \partial^2_rR(r)+\left(\frac{3}{r}+\frac{f'(r)}{f(r)}\right)\partial_rR(r)+ $$
 \begin{equation}
 \frac{1}{f(r)}\left(\frac{\omega^2}{f(r)}-\frac{\kappa^2}{r^2}-m^2\right)R(r)=0~,
 \label{radial}
\end{equation}%
where $-\kappa^2=-\ell(\ell+2)$ is the eigenvalue of the Laplacian in the base submanifold. Now, defining $R(r)$ as
$ R(r)=F(r) r^{-3/2} $
and using the tortoise coordinate $r_*$ given by
$ dr_* = dr/f(r) $,
 the Klein-Gordon equation can be written as a one-dimensional Schr\"{o}dinger equation,
 \begin{equation}\label{ggg}
 -\frac{d^{2}F(r_*)}{dr_*^{2}}+V(r)F(r_*)=\omega^{2}F(r_*)\,,
 \end{equation}
 with an effective potential $V(r)$, which is parametrically thought of as $V(r_*)$, given by
  \begin{equation}\label{pot}
 V(r)=\frac{f(r)}{r^2} \left(\frac{3}{4}f(r)+\frac{3}{2}r\frac{df}{dr}+\kappa^2+m^2 r^2 \right)~.
 \end{equation}
The effective potential
diverges at spatial infinity, and one can check that it is positive definite everywhere outside the event horizon.

\begin{figure}
\resizebox{.8\linewidth}{!}{\includegraphics*{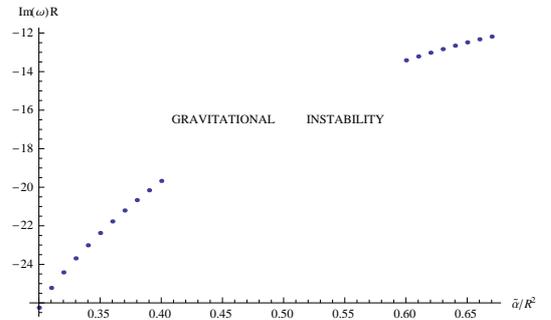}}
\caption{The nonperturbative (in $\alpha$) purely imaginary modes of a scalar field for $D=6$, $\ell =1$, $r_{H}/R =5$ are growing when $\alpha$ is decreasing. The region in the middle has damped frequencies for a test scalar field but corresponds to eikonal instability of gravitational perturbations of the background.}\label{non-perturbative}
\end{figure}

\begin{figure}
\resizebox{.9\linewidth}{!}{\includegraphics*{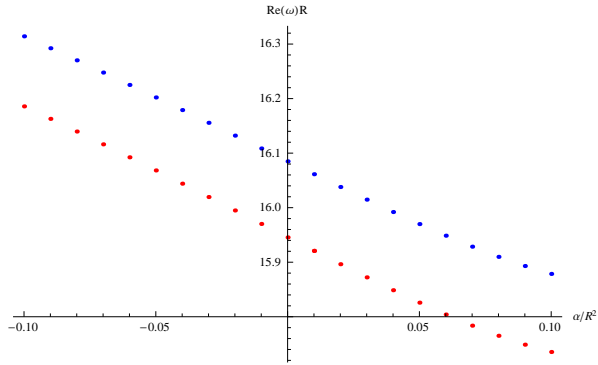}}
\caption{Real part of quasinormal modes (perturbative in $\alpha$ branch) of a scalar field for $D=5$, $\ell =0$ (lower) and $1$ (upper), $r_{H}/R =5$.  The mode on the ordinate axis represents the Schwarzschild-AdS modes: $\omega = 15.9454 - 13.6914 i$  ($\ell=0$) and $\omega = 16.0849 - 13.6487 i$ ($\ell=1$).}\label{Perturbative1}
\end{figure}

\begin{figure}
\resizebox{.9\linewidth}{!}{\includegraphics*{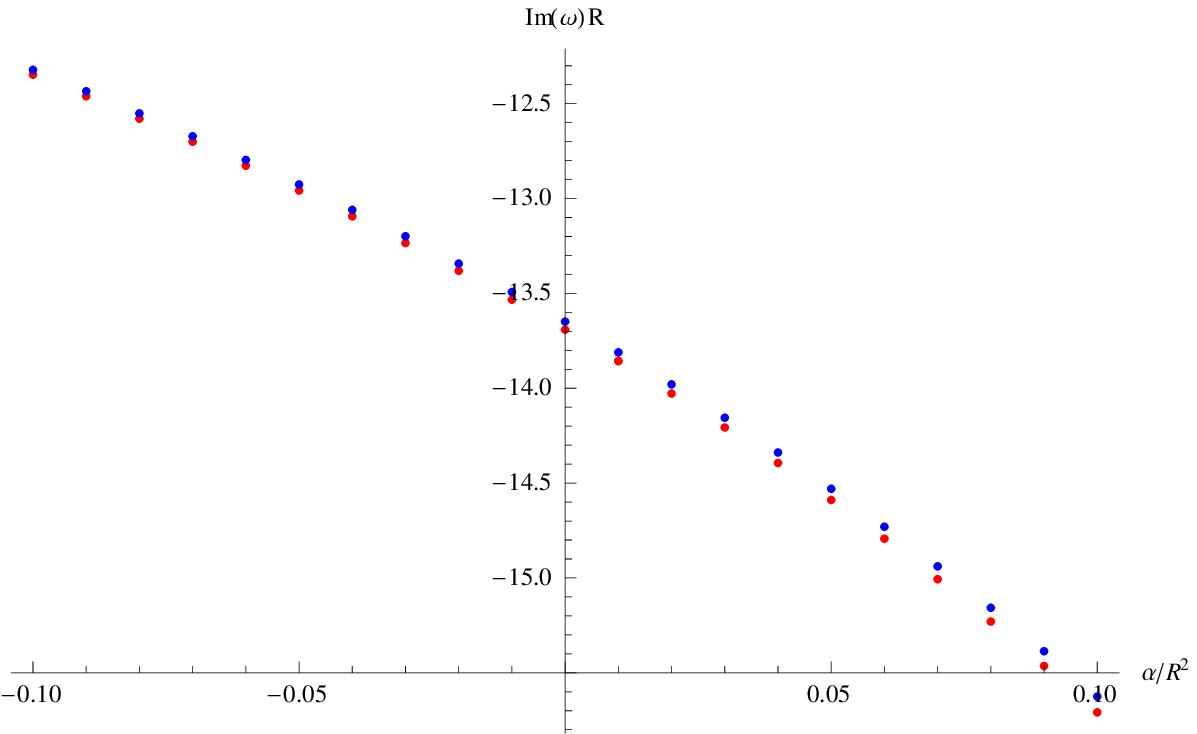}}
\caption{Imaginary part of quasinormal modes (branch perturbative in $\alpha$) of a scalar field for $D=5$, $\ell =0$ (red, lower) and $1$ (blue, upper), $r_{H}/R =5$. The mode on the ordinate axis represents the Schwarzschild-AdS modes: $\omega = 15.9454 - 13.6914 i$ ($\ell=0$) and $\omega = 16.0849 - 13.6487 i $ ($\ell=1$).}\label{Perturbative2}
\end{figure}

The numerical search of quasinormal modes of a test scalar field in the background of a black hole is motivated only inside the range of black hole parameters, which guarantees the stability against gravitational perturbations \cite{Konoplya:2017ymp}. Thus, from Fig. 1 in Ref. \cite{Konoplya:2017ymp}, one can see that for $r_{H}/R = 5$, $D =5$ the AdS black hole is stable at
 $-0.1 \lesssim (\alpha/R^2) \lesssim 0.12$ and for $D=6$ it is stable at $ -0.06 \lesssim (\alpha/R^2) \lesssim 0.14$ and $0.2 \lesssim (\alpha/R^2) \lesssim 0.33$. Therefore, the quasinormal modes obtained here for a test scalar field by the shooting method are given only within the above stability region (see Figs. \ref{non-perturbative}, \ref{Perturbative1}, and \ref{Perturbative2}).

\begin{figure}
\resizebox{.9\linewidth}{!}{\includegraphics*{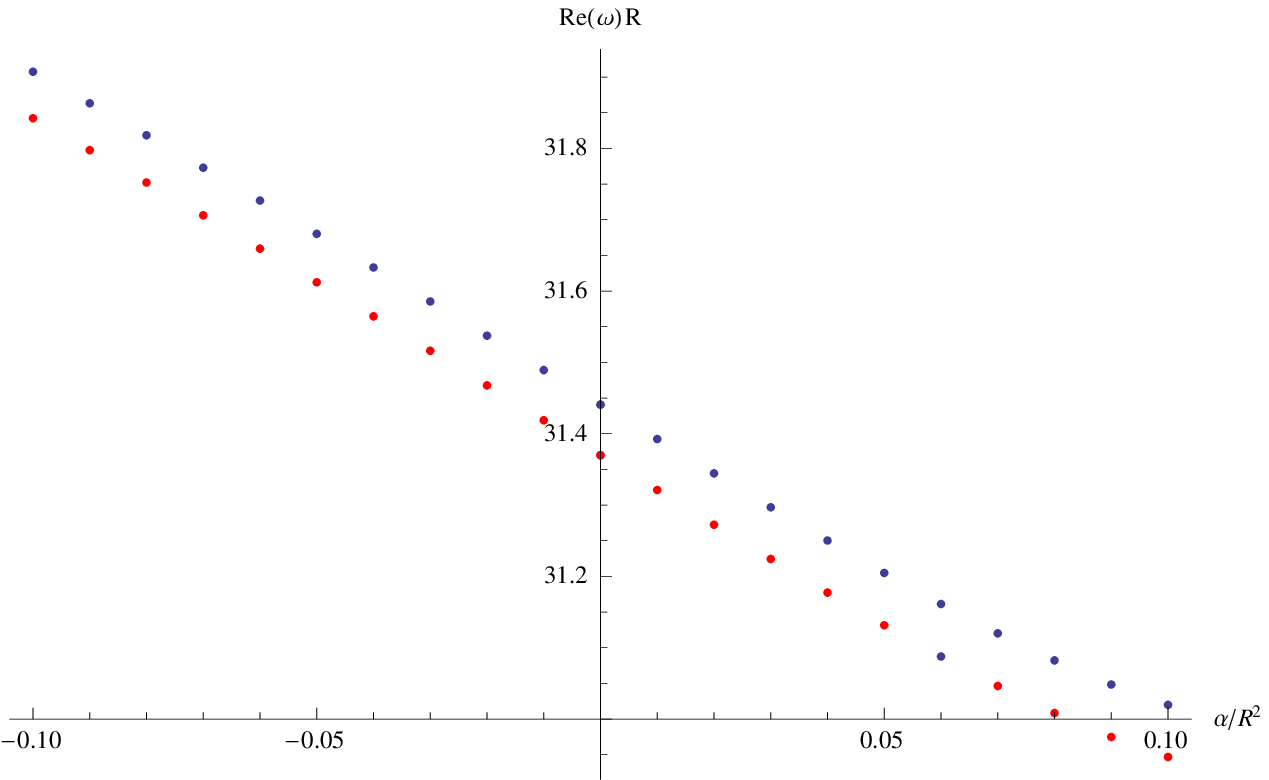}}
\caption{Real part of quasinormal modes (branch perturbative in $\alpha$) of a scalar field for $D=5$, $\ell =0$ (lower) and $1$ (upper), $r_{H}/R =10$.  The mode on the ordinate axis represents the Schwarzschild-AdS modes: $\omega = 31.3699 - 27.4457 i $  ($\ell=0$) and $\omega = 31.4408 - 27.4242  i$ ($\ell=1$).}\label{Perturbative3}
\end{figure}

\begin{figure}
\resizebox{.9\linewidth}{!}{\includegraphics*{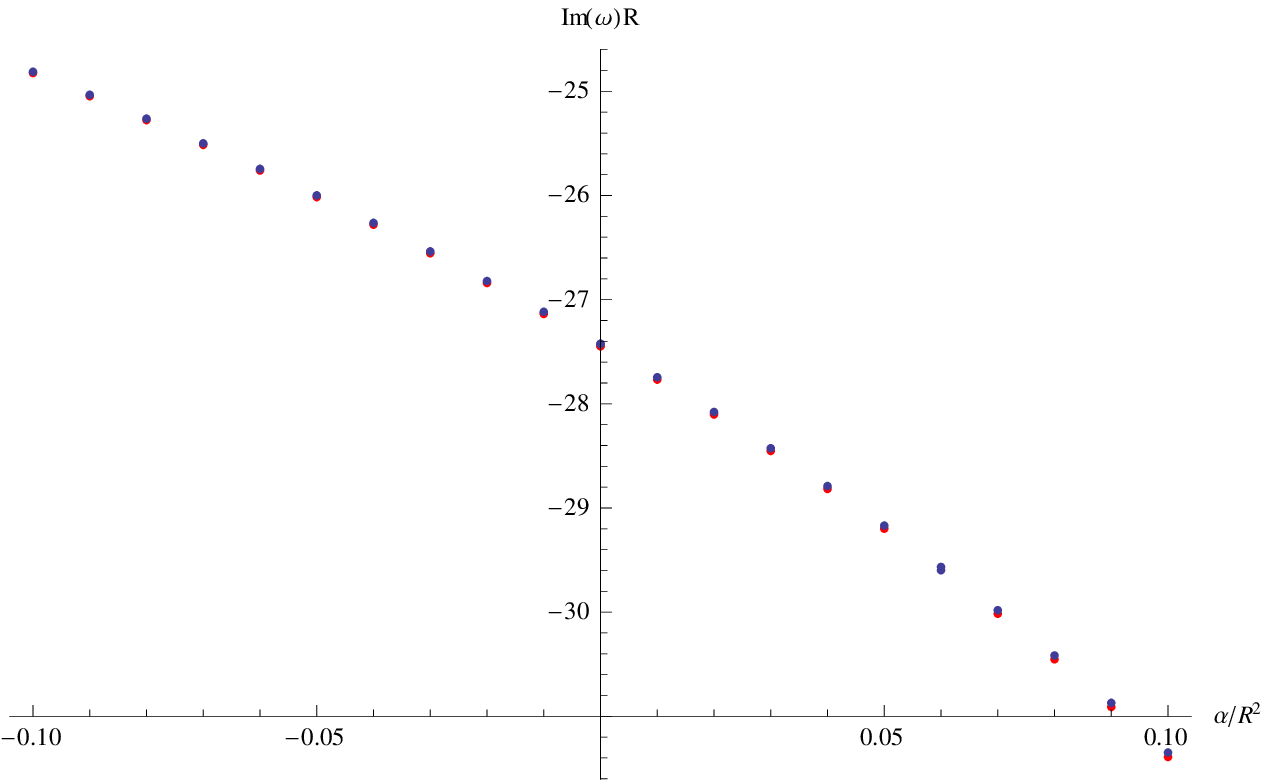}}
\caption{Imaginary part of quasinormal modes (branch perturbative in $\alpha$) of a scalar field for $D=5$, $\ell =0$ (red, lower) and $1$ (blue, upper), $r_{H}/R =10$. The mode on the ordinate axis represents the Schwarzschild-AdS modes: $\omega = 31.3699 - 27.4457 i$ ($\ell=0$) and $\omega = 31.4408 - 27.4242 i $ ($\ell=1$).}\label{Perturbative4}
\end{figure}

From Figs. \ref{Perturbative1}--\ref{Perturbative4} we can see that for $D=5$ the real oscillation frequencies obey the following fits:
\begin{equation}\nonumber
Re (\omega) \cdot R = 31.3763 (1 - 0.147379 (\aa/R^2) ) \quad (\ell =0)
\end{equation}
\begin{equation}\nonumber
Re (\omega) \cdot R= 31.4427 (1 - 0.147769 (\aa/R^2) ) \quad (\ell=1).
\end{equation}
Hence, we see that $Re (\omega)$ for the Schwazrschild-AdS black hole and its Gauss--Bonnet generalization are related as
\begin{equation}\label{perturbative-mode}
Re (\omega_{GB})  = Re (\omega_{SAdS}) (1+ K(D) (\aa/R^2))\, ,
\end{equation}
where $K(5) \approx 0.15$ for $D=5$.
Thus, at least for small $\alpha$, the branch of the spectrum perturbative in $\alpha$ has oscillation frequencies linear in $\alpha$.
The damping rates, given by the imaginary part of $\omega$, depend weakly on the multipole number $\ell$ and increase when $\alpha$ is growing, as can be seen from the almost coinciding curves in Figs. \ref{Perturbative2} and \ref{Perturbative4}.

From Fig. \ref{non-perturbative} one can see that there is also another branch of purely imaginary quasinormal modes which increase when $\alpha$ is decreasing. Therefore, when moving along the nonperturbative branch of the mode, one cannot reach the limit $\alpha \rightarrow 0 $ numerically. Similar purely imaginary modes were found for gravitational perturbations of the planar AdS black holes in higher curvature corrected theories \cite{Grozdanov:2016fkt} and of asymptotically flat Gauss-Bonnet black holes \cite{Konoplya:2008ix}.

Let us now compare the obtained exact solutions for $\omega$ with the results of numerical computations done via the shooting method. The shooting method is based on numerical integration of the wave equation from the event horizon and matching it with the proper asymptotic expansion at infinity. Perturbations satisfy the Dirichlet boundary conditions at infinity and the requirement of the purely ingoing waves onto the black hole event horizon \cite{Konoplya:2017ymp}.

We were unable to reproduce accurately the analytical results when $r_{H} \lesssim R \sqrt{\ell (\ell +2)}$ with the shooting method (see Table \ref{Table1}) because the whole procedure becomes unstable and apparently requires the improvement of accuracy. In principle, the shooting method is based on the convergent procedure and must be as accurate as one wishes, provided the integration is performed properly and with controlled accuracy. Practically, integrating the differential equations and fitting the results of integration with the required asymptotic behavior with the help of {\it Mathematica} built-in functions, we were unable to achieve the desired accuracy.

The Schwarzschild--Gauss--Bonnet-AdS black hole considered here has imaginary modes given by Eq. (\ref{w1}) at $\alpha = R^2/2$.
For values of $\alpha$ near $R^2/2$, the damping rate of the purely imaginary scalar modes is increasing, when $\alpha$ decreases. Thus, the $\omega$ analytically found in (\ref{w1}) goes over into the new nonperturbative quasinormal modes, when $\alpha$ is small enough to guarantee gravitational stability. As nonperturbative modes exist also in $R^4$ theory \cite{Grozdanov:2016vgg} and for asymptotically flat Gauss-Bonnet black holes \cite{Konoplya:2008ix}, we suppose that this phenomenon might be sufficiently general and independent of the number of spacetime dimensions, black hole asymptotic behavior, or the spin of a field.

\begin{table}
  \centering
  \begin{tabular}{|c|c|c|}
  \toprule[1pt]\midrule[0.3pt]
 $\ell$  &  Exact  & Numerical \\
      \hline
1  &  $- 12.25543735 i$ & $- 12.25543726 i$ \\
     \hline
2  &  $- 12.70849738 i$ & $- 12.70849733 i$ \\
     \hline
3  &  $- 13.41742431 i$ & $- 13.41742435 i$ \\
     \hline
4  &  $- 14.53589838 i$ & $- 14.53589752 i$ \\
     \hline
5  & $ - 17 i$ & $-  17.043 i$ \\
     \hline
6  &  $3.4641 - 18. i$ & $3.2833 - 17.3917 i$ \\

  \midrule[0.3pt]\bottomrule[1pt]
   \end{tabular}
  \caption{Comparison of the exact $\omega$ for $m=n=0$, $r_H = 6 R$ with the results found by the shooting method for various values of the multipole number $\ell$, $D=5$. The last two values of $\ell =5, 6$ have bad convergence of the numerical procedure.}\label{Table1}
\end{table}

\section{Exact solutions}

In this section, we deduce the exact analytic solution for the master equation (\ref{ggg}) in the background given by (\ref{metric}).
First, we shall consider a test scalar field equation. The background (\ref{metric}) is known to be unstable under gravitational perturbations, and the instability is eikonal; that is, it develops at higher $\ell$ and is accompanied by the breakdown of the well-posedness of the initial value problem because of the absence of convergence of a signal when summing over different $\ell$.
Therefore, strictly speaking, any frequencies found for this case cannot represent quasinormal modes of any real black hole. However,
\emph{the perturbation with fixed $\ell$}, formally, does not have any problem with the initial conditions, and we can solve the corresponding wave equation in order to be able to see the effect of \emph{$\ell$ divergence} analytically and to check the correctness and accuracy of our numerical (shooting) computations by the independent and analytical calculations.

\subsection{Scalar field}

Under the change of variable $u=1-\frac{r_H^2}{r^2}$, the Klein--Gordon equation (\ref{radial})  can be written as
$$ \partial^2_uR(u)+\frac{1}{u(1-u)}\partial_uR(u)+ $$
\begin{equation}
\frac{1}{u(1-u)}\left(\frac{\omega^2 R^4}{4ur_H^{2}}-\frac{\kappa^2R^2}{4r_H^2}-\frac{m^2 R^2}{4(1-u)}\right)R(u)=0~,
\end{equation}
and if, in addition, we define $R(u)=u^{\alpha}(1-u)^\beta K(u)$, the above equation leads to the hypergeometric equation
\begin{equation}\label{hypergeometric}
 u(1-u)K''(u)+\left[c-(1+a+b)u\right]K'(u)-ab K(u)=0~,
\end{equation}
 where
\begin{equation}
\alpha =  \pm\frac{i\omega R^2}{2r_H}, \quad \beta= \frac{1}{2}\left(2\pm \sqrt{4+m^2R^2}\right)
\end{equation}
and the constants are given by
\begin{equation}\label{a}
a_{1,2}= \alpha +\beta -\frac{1}{2}\pm \frac{\sqrt{r_H^2-\kappa^2R^2}}{2 r_H},
\end{equation}
\begin{equation}
b_{1,2}= \alpha +\beta -\frac{1}{2}\mp \frac{\sqrt{r_H^2-\kappa^2R^2}}{2 r_H}, \quad c=1+2\alpha.
\end{equation}
The general solution of the hypergeometric equation~(\ref{hypergeometric}) is
$$ K(u)= C_{1}\,\, {_2}F{_1}(a,b,c;u)+ $$
\begin{equation}
\label{HSolution}
C_2u^{1-c}\,{_2}F{_1}(a-c+1,b-c+1,2-c;u)~,
\end{equation}
and it has three regular singular points at $u=0$, $u=1$, and
$u=\infty$. Here, ${_2}F{_1}(a,b,c;u)$ is a hypergeometric function,
and $C_{1}$ and $C_{2}$ are integration constants.
Thus, in the vicinity of the horizon $u=0$ and using
the property $F(a,b,c,0)=1$, the function $R(u)$ behaves as
\begin{equation}\label{Rhorizon}
R(u)=C_1 e^{\alpha \ln u}+C_2 e^{-\alpha \ln u},
\end{equation}
so that the scalar field $\psi$, for $\alpha=\alpha_-$, can be written as follows:
\begin{equation}
\psi\sim C_1 e^{-i\omega (t+ R^2\ln u /(2r_H))}+C_2
e^{-i\omega (t-R^2\ln u/(2r_H))}~.
\end{equation}

\emph{Dirichlet boundary condition.} Here, the first term represents an ingoing wave, and the second represents an outgoing wave near the black hole horizon. Imposing the requirement of
only ingoing waves on the event horizon, we fix $C_2=0$. Then, the radial solution can be written as
$$ R(u)=C_1 e^{\alpha \ln u}(1-u)^\beta {_2}F{_1}(a,b,c;u)= $$
\begin{equation}\label{horizonsolution}
C_1e^{-\frac{i\omega R^2}{2r_H} \ln u}(1-u)^\beta{_2}F{_1}(a,b,c;u)~.
\end{equation}
To implement boundary conditions at infinity ($u=1$), we
apply Kummer's formula
for the hypergeometric function \cite{M. Abramowitz},
$$ {_2}F{_1}(a,b,c;u)=\frac{\Gamma(c)\Gamma(c-a-b)}{\Gamma(c-a)\Gamma(c-b)}F_1+ $$
\begin{equation}\label{relation}
(1-u)^{c-a-b}\frac{\Gamma(c)\Gamma(a+b-c)}{\Gamma(a)\Gamma(b)}F_2,
\end{equation}
where
\begin{equation}
F_1={_2}F{_1}(a,b,a+b-c,1-u)~,
\end{equation}
\begin{equation}
F_2={_2}F{_1}(c-a,c-b,c-a-b+1,1-u)~.
\end{equation}
Taking into consideration the above expression, the radial function~(\ref{horizonsolution}) reads
$$ R(u) = C_1 e^{-\frac{i\omega R^2}{2r_H} \ln u}(1-u)^\beta\frac{\Gamma(c)\Gamma(c-a-b)}{\Gamma(c-a)\Gamma(c-b)} F_1 + $$
\begin{equation}
C_1 e^{-\frac{i\omega R^2}{2r_H}  \ln u}(1-u)^{2-\beta}\frac{\Gamma(c)\Gamma(a+b-c)}{\Gamma(a)\Gamma(b)}F_2,
\end{equation}
and at infinity, it can be written as
$$ R_{asymp.}(u) = C_1 (1-u)^\beta \frac{\Gamma(c)\Gamma(c-a-b)}{\Gamma(c-a)\Gamma(c-b)}+ $$
\begin{equation}\label{R2}\
C_1 (1-u)^{2-\beta}\frac{\Gamma(c)\Gamma(a+b-c)}{\Gamma(a)\Gamma(b)}~.
\end{equation}
Thus, the field at infinity vanishes
if $a=-n$ or $b=-n$ for $n=0,1,2,...$.
Therefore, the discrete frequencies for the $D=5$ Einstein--Gauss--Bonnet-AdS black hole at $\alpha = R^2/2$ are given by
\begin{equation}\label{w1}
\omega_1 = -\frac{i}{R^2}\left(1+2n+\sqrt{4+m^2 R^2}\right)r_H \pm \frac{i}{R^2}\sqrt{r_H^2-\kappa^2R^2}~.
\end{equation}
The imaginary part of the quasinormal frequencies (QNFs) is always negative, so the propagation of a scalar field is formally stable in this background. When the multipole number $\ell$ is large enough in comparison with the black hole size, then the second term acquires the nonzero  real part. In other words, for sufficiently high $\ell$ or sufficiently small black hole radius,
$r_{H} < R \sqrt{\ell (\ell +2)}$,
the purely imaginary frequencies become oscillating, so for black holes with a radius that is smaller than $\sqrt{3} R$, only the $s$ wave ($\ell =0$) has purely imaginary modes.


\emph{Neumann boundary conditions.}
It is also possible to consider that the flux of the scalar field vanishes at infinity, which implies the Neumann boundary conditions. The Dirichlet boundary condition also leads to discrete frequencies for $m^2 > 0$ but not for $m^2 < 0$.
Indeed, in the range $-4<m^2R^2<0$, then $1<\beta_{+}<2$ and $0<\beta_{-}<1$, and at spatial infinity ($u \rightarrow 1$) $R_{asymp.}\rightarrow 0$ [Eq. 32], which leads to the continuous spectrum.
Thus, the Neumann boundary conditions allow one also to describe tachyons within the supergravity context \cite{Breitenlohner:1982bm}.

To consider Neumann boundary condition at infinity, $u\rightarrow 1$, the flux
\begin{equation}
F=\frac{\sqrt{-g}g^{rr}}{2i}\left( R^{\ast }\partial _{r}R-R\partial
_{r}R^{\ast }\right)~,
\end{equation}
at infinity, is given by
$$ F\left( u\rightarrow 1\right)  =  $$
\begin{equation}\nonumber
\begin{split}
\frac{2r_H^4}{R^2} \left\vert C_{1}\right\vert ^{2}  Im  \Big( \alpha  \left\vert A\right\vert ^{2} (1-u)^{2\beta-1}+\alpha  \left\vert B \right\vert ^{2} (1-u)^{3-2\beta}  +  \\
\frac{ab}{c} A^\prime A^*(1-u)^{2\beta-1}+\frac{ab}{c} B^\prime B^*(1-u)^{2-2\beta}+ \frac{ab}{c} A^*B^\prime \Big)~,
\end{split}
\end{equation}
where
\begin{equation}
A=\frac{\Gamma \left( c\right) \Gamma \left( c-a-b\right) }{\Gamma
\left( c-a\right) \Gamma \left( c-b\right) }~,
\end{equation}
\begin{equation}
B=\frac{\Gamma \left( c\right) \Gamma \left( a+b-c\right) }{\Gamma
\left( a\right) \Gamma \left( b\right) }~,
\end{equation}
$A^\prime = \frac{c}{c-a-b-1}A$ and $B^\prime = \frac{c(a+b-c)}{ab}B$. For, $-4<m^{2}R^2<0$,
$1<\beta _{+}<2$ and $0<\beta _{-}<1$. So, for $\beta=\beta_+$, the flux vanishes at infinity if $a=-n$ or $b=-n$, which leads to the same quasinormal modes that we have found by imposing the Dirichlet boundary condition. Also, for $\beta=\beta_-$ and $0<\beta _{-}<1/2$, the flux vanishes at infinity if $c-a=-n$ or $c-b=-n$, giving the same quasinormal modes that we have found by imposing the Dirichlet boundary condition. However, for $1/2<\beta _{-}<1$ $(-4<m^2R^2<-3)$ the flux vanishes at infinity if $c-a=-n$ or $c-b=-n$ and $a=-n$ or $b=-n$. Then, it is possible to obtain a new set of quasinormal modes, which is
\begin{equation}\label{w2}
\omega_2 = -\frac{i}{R^2}\left(1+2n-\sqrt{4+m^2 R^2}\right)r_H \pm \frac{i}{R^2}\sqrt{r_H^2-\kappa^2R^2}~.
\end{equation}
The new set of frequencies presents a negative  imaginary part in some cases. Note that for some cases $\omega$ has a real part, and for $\ell=0$ and $n=0$, $\omega$ is purely imaginary and positive.
Thus, the scalar field at $\ell=0$ has the fundamental mode ($n=0$)
\begin{equation}
\omega_2=\frac{2i}{R^2}\left(\sqrt{1+\frac{m^2 R^2}{4}} \right)r_H,
\end{equation}
which means instability even of a scalar field under the Neumann boundary conditions for both tachyons and tardyons. The latter relation means also the absence of the analog of the Breitenlohner--Freedman gap of stability for small negative $m^2$ under Neumann boundary conditions.
Although the case $\alpha = R^2/2$ considered here is unphysical, this instability of a fixed-$\ell $ perturbation may indicate  also  possible instability at small $\alpha$, that is, when the black hole background itself is stable against gravitational perturbations.

\subsection{Gravitational perturbations}


The gravitational perturbations can be treated separately for scalars, vectors, and tensors
relatively the rotation group on the $(D-2)$-dimensional sphere and then treated independently from each other. The explicit expressions for the effective potentials can be found in Refs. \cite{Gleiser:2005ra} and \cite{Dotti:2005sq}.

\emph{Scalar type of gravitational perturbations}. The effective potential for the scalar type of \emph{gravitational} perturbations of the metric (\ref{metric}) has the form
\begin{equation}
V_s(r)=\frac{(r^2-r_H^2)(35r^2-15r_{H}^2-4R^2 \ell (\ell+2))}{4r^2R^4}.
\end{equation}

In terms of the tortoise coordinate
$r_{\ast}=-(R^2/r_H) \text{arccoth}(r/r_H)$,
the Schwarzschild-like coordinate $r$  can be expressed as
$r=-r_H \coth (r_H r_{\ast}/R^2)$,
while the effective potential is given by
\begin{equation}\nonumber
V_s(r_{\ast})=\frac{35 r_H^2}{4 R^4 \sinh^2\left( \frac{r_H r_{\ast}}{R^2} \right)}-  \frac{15 r_H^2 +8R^2 \ell+4 R^2 \ell^2}{4 R^4 \cosh^2 \left( \frac{r_H r_{\ast}}{R^2} \right)} \,.
\end{equation}

Now, changing the variable $x=\cosh^{-2} \left( \frac{r_H r_{\ast}}{R^2} \right)$, the Schr\"{o}dinger equation (\ref{ggg}) can be written as
$$ x(1-x)\frac{d^2 F}{dx^2}+(1-\frac{3}{2}x) \frac{dF}{dx}+ $$
\begin{equation}
\left( \frac{R^4 \omega^2}{4 r_H^2 x}-\frac{35}{16(1-x)}+\frac{15 r_H^2+8R^2\ell+4R^2 \ell^2}{16 r_H^2} \right) F=0 \,,
\end{equation}
and if, in addition, we define $F(x)=x^{\alpha}(1-x)^\beta K(x)$, the above equation leads to the hypergeometric equation (\ref{hypergeometric}),
 where
\begin{equation}
\alpha =  \pm\frac{i\omega R^2}{2r_H}, \quad \beta_+=\frac{7}{4}, \quad \beta_-=-\frac{5}{4},
\end{equation}
and the constants are given by
\begin{equation}
a_{1,2}= \alpha +\beta +\frac{1}{4}\pm \frac{\sqrt{4r_H^2+\ell (\ell+2)R^2}}{2 r_H}~,
\end{equation}
\begin{equation}
b_{1,2}= \alpha +\beta +\frac{1}{4}\mp \frac{\sqrt{4r_H^2+\ell (\ell+2)R^2}}{2 r_H}~,
\end{equation}
\begin{equation}
c=1+2\alpha~.
\end{equation}
From now and on, we shall consider the cases $\alpha=\alpha_-$ and $\beta=\beta_+$. Therefore,  in the vicinity of the horizon $x=0$,
the function $R(x)$ behaves as
\begin{equation}
R(x)=C_1 e^{\alpha \ln x}+C_2 e^{-\alpha \ln x},
\end{equation}
so the perturbation $\psi$ for $\alpha=\alpha_-$ can be written as follows:
\begin{equation}\nonumber
\psi\sim C_1 e^{-i\omega (t+ R^2\ln x /(2r_H))}+C_2
e^{-i\omega (t-R^2\ln x/(2r_H))}~.
\end{equation}

Here, the first term represents an ingoing wave, and the second represents an outgoing wave near the black hole horizon. Imposing the requirement of
only ingoing waves on the event horizon, we fix $C_2=0$. Then, the radial
solution can be written as
$$ F(x)=C_1 e^{\alpha \ln x}(1-x)^\beta {_2}F{_1}(a,b,c;x)= $$
\begin{equation}\label{horizonsolutions}
C_1e^{-\frac{i\omega R^2}{2r_H} \ln x}(1-x)^\beta{_2}F{_1}(a,b,c;x)~.
\end{equation}
To implement boundary conditions at infinity ($x=1$), we
apply Kummer's formula (\ref{relation})
for the hypergeometric function \cite{M. Abramowitz},
Thus, the radial function~(\ref{horizonsolutions}) reads
\begin{eqnarray}
\nonumber F(x) &=& C_1 e^{-\frac{i\omega R^2}{2r_H} \ln x}(1-x)^\beta\frac{\Gamma(c)\Gamma(c-a-b)}{\Gamma(c-a)\Gamma(c-b)} F_1\\
\nonumber &&+C_1 e^{-\frac{i\omega R^2}{2r_H}  \ln
x}(1-x)^{1/2-\beta}\frac{\Gamma(c)\Gamma(a+b-c)}{\Gamma(a)\Gamma(b)}F_2~, \\
\end{eqnarray}
and at infinity, it can be written as
$$ F_{asymp.}(x) = C_1 (1-x)^\beta \frac{\Gamma(c)\Gamma(c-a-b)}{\Gamma(c-a)\Gamma(c-b)}+ $$
\begin{equation}\label{asym}
C_1 (1-x)^{1/2-\beta}\frac{\Gamma(c)\Gamma(a+b-c)}{\Gamma(a)\Gamma(b)}~.
\end{equation}
Thus, the field at infinity vanishes
if $a=-n$ or $b=-n$ for $n=0,1,2,...$.
Therefore, the frequencies are given by
\begin{equation}
\omega = -\frac{i}{R^2}\left(2r_H (2 +n) \pm \sqrt{\ell R^2 (2 + \ell) +4 r_H^2}\right).
\end{equation}

From the above formula, we can see that when $\ell$ is large $Im (\omega)$ becomes positive, so an eikonal instability develops. Similar analytic formulas for the instability of Gauss--Bonnet-AdS black branes were found in Refs. \cite{Grozdanov:2016fkt} and \cite{Grozdanov:2016vgg}. Thus, from here and Ref. \cite{Konoplya:2017ymp}, we see that the parametric region $\alpha \lesssim R^2/2$ discussed in the context of the possible holographic description of the quantum dissipationless liquids \cite{Brigante:2007nu} lies well inside the region of instability.

\emph{Vector type of gravitational perturbations.} For vector gravitational perturbations, the effective potential is given by
\begin{equation}
V_v=\frac{5(3r^4-10r^2r_H^2+7r_H^4)}{4R^4r^2} \,,
\end{equation}
and in terms of the tortoise coordinate,
\begin{equation}
V_v(r_{\ast})=\frac{15 r_H^2}{4 R^4 \sinh^2\left( \frac{r_H r_{\ast}}{R^2} \right)}-\frac{35 r_H^2}{4 R^4 \cosh^2 \left( \frac{r_H r_{\ast}}{R^2} \right)} \,.
\end{equation}
The Schr\"{o}dinger equation, under the change of variable $x=\cosh^{-2} \left( \frac{r_H r_{\ast}}{R^2} \right)$ becomes
$$ x(1-x)\frac{d^2 F}{dx^2}+(1-\frac{3}{2}x) \frac{dF}{dx}+ $$
\begin{equation}
\left( \frac{R^4 \omega^2}{4 r_H^2 x}-\frac{15}{16(1-x)}+\frac{35}{16} \right) F=0 \,.
\end{equation}
Defining $F(x)=x^{\alpha}(1-x)^\beta K(x)$, the above equation leads to the hypergeometric equation (\ref{hypergeometric}),
 where
\begin{equation}
\alpha =  \pm\frac{i\omega R^2}{2r_H} \quad \beta_+=\frac{5}{4}, \quad \beta_-=-\frac{3}{4},
\end{equation}
and the constants are given by
\begin{equation}
a_{1}= \alpha +\beta -\frac{5}{4}~, \,\,\, a_{2}= \alpha +\beta +\frac{7}{4}~,
\end{equation}
\begin{equation}
b_{1}= \alpha +\beta +\frac{7}{4}, \,\,\, b_{2}= \alpha +\beta -\frac{5}{4}, ~ c=1+2\alpha.
\end{equation}
Following the above procedure, we can obtain expression (\ref{asym}) so that the field at infinity vanishes
if $a=-n$ or $b=-n$ for $n=0,1,2,...$.
Therefore, the frequencies are given by
\begin{equation}
\omega_1 = -\frac{2ir_Hn}{R^2}, \,\,\, \omega_2 = -\frac{2ir_H(3+n)}{R^2}.
\end{equation}

\emph{Tensor type of gravitational perturbations.}
The effective potential for the tensor perturbations has the form
\begin{equation}
V_t=\frac{(r^2-r_H^2)(35r^2-15r_{H}^2+12R^2 \ell (\ell+2))}{4r^2R^4}
\end{equation}
Acting in a similar fashion with the scalar type of gravitational perturbations and by using the same changes of variables, we can find the exact solution for $\omega$,
\begin{equation}\nonumber
\omega = -\frac{i}{R^2}\left(2r_H (2 +n) \pm \sqrt{-3 \ell R^2 (\ell+2) +4 r_H^2}\right).
\end{equation}

\section{Final remarks}
\label{conclusion}

Here, we have investigated the quasinormal spectrum of a scalar field in the background of the $D=5, 6$ Einstein--Gauss--Bonnet-AdS black holes.
(Higher $D$ require higher than second curvature corrections to the Einsteinian action.)
We have shown that the quasinormal spectrum consists of two different branches. One of them has an Einsteinian limit when $\alpha \rightarrow 0$, while the other consists from purely imaginary modes of which the damping rate is increasing, when $\alpha$ decreases. This branch is, thereby, nonperturbative in $\alpha$. Previously, purely imaginary modes in higher than four dimensions were found in the gravitational spectra of Gauss-Bonnet black holes \cite{Konoplya:2017ymp} and branes \cite{Grozdanov:2016fkt}, and in the vector type of gravitational perturbations in the Schwarzschild-AdS solution \cite{Konoplya:2003dd}, but never for the test scalar field.

At a fixed Gauss-Bonnet coupling $\alpha =R^2/2$, we have found exact solution of the master perturbation equations. Although this case suffers from the eikonal instability \cite{Konoplya:2017ymp}, and thereby from the absence of convergence in $\ell$, the fixed $\ell$ perturbations do not have such a problem, and this allowed us to check the numerical calculations by analytical expressions for $\omega$. We have shown that the shooting method has limitations when searching frequencies at large multipole numbers $\ell$ or high overtones $n$.

This work could be extended in a number of ways. First of all, the influence of corrections of higher than the second order in curvature can be analyzed, and this way, higher than $D=6$ spacetimes can be included in the consideration self-consistently. Thus, it would also be interesting to consider quasinormal modes of tachyons at the Neumann boundary conditions for gravitationally stable Gauss-Bonnet black holes.

\acknowledgments

R.~K. would like to thank A. Zhidenko and A. Starinets for useful discussions and the Bridging Grant of the University of T\"ubingen for support. This work was also partially funded by the Comisi\'{o}n Nacional de Ciencias y Tecnolog\'{i}a through FONDECYT Grant No. 11140674 (P. A. G.) and by the Direcci\'{o}n de Investigaci\'{o}n y Desarrollo de la Universidad de La Serena (Y.V.). P. A. G. acknowledges the hospitality of the Universidad de La Serena where part of this work was undertaken.

\end{document}